\documentclass[aps,twocolumn,a4paper,10pt]{revtex4-1}
\usepackage{mathrsfs}
\usepackage{graphicx}
\usepackage{latexsym}
\usepackage{amsmath}
\usepackage{amssymb}
\usepackage{textcomp}
\usepackage{amsbsy}
\usepackage{graphics}
\usepackage{epstopdf}
\usepackage{color}

\begin{document}
\title{Holographic correspondence of $F(R)$ gravity with/without matter fields}
\author{Tanmoy Paul}
\email{pul.tnmy9@gmail.com}
\affiliation{School of Physical Sciences,\\
Indian Association for the Cultivation of Science,\\
2A $\&$ 2B Raja S.C. Mullick Road, Kolkata - 700 032, India.\\}

\begin{abstract}
In this paper, we apply the holographic principle at the early universe, obtaining an inflation realization of holographic origin. First we show that 
under the consideration of extended infrared cut-offs, there exist a holographic correspondence of various F(R) (such as quadratic, cubic, exponential 
F(R) models) cosmological models in absence of matter fields. Then we establish that such holographic correspondence is not only confined 
to the vacuum F(R) gravity but also can be extended to the higher curvature models along with matter fields. In presence of matter field, in particular 
the second rank antisymmetric Kalb-Ramond field, all the considered F(R) models lead to a viable inflationary phenomenology consistent with 
Planck 2018 constraints. We further extend our discussion of holographic correspondence to F(T) cosmological models.
\end{abstract}

\maketitle

\section{Introduction}
The holographic principle originates from black hole thermodynamics and string theory and establishes a connection of the Ultraviolet cutoff 
of a quantum field theory, which is related to the vacuum energy, with the largest distance of this theory \cite{1,2,3}. This consideration has been 
applied extensively at a cosmological framework at late times, in which case the obtained vacuum energy constitutes 
a dark energy sector of holographic origin, called holographic dark energy \cite{6}. At this stage it deserves mentioning that 
the most general holographic dark energy is given by the one with Nojiri-Odintsov cut-off \cite{7} 
and it is interesting that it may be applied to covariant 
theories too \cite{nojiri_new,martiros}. 
In particular, the holographic energy density is proportional to the inverse squared Infrared cutoff $L_{IR}$, namely 
$\rho_h = \frac{3}{\kappa^2L_{IR}^2}$, with $\kappa^2$ the gravitational constant. Therefore the holographic 
Friedmann equation can be written as $H = \sqrt{\kappa^2\rho_h/3} = 1/L_{IR}$. Holographic dark energy acquires a vast 
phenomenology, both in its basic \cite{6,7,8,9} as well as in its various extensions \cite{15,16,17,18}, and it can also fit well with 
observations \cite{23,24,25,26}. Beside the application of holographic principle in late time cosmology and dark energy epoch, 
recently the holographic principle 
is also applied to early universe cosmology to realize the inflation from holographic origin \cite{odintsov}. In the case of early universe 
cosmology, holographic inflation proves to have a very interesting phenomenology, and thus it becomes a 
good candidate for the description of the early universe. However at the end of \cite{odintsov}, the authors clearly showed that there exist a holographic 
correspondence of Starobinsky $R^2$ gravity inflation in absence of matter fields i.e the cosmological field equations of vacuum $R^2$ gravity 
can be reproduced from the holographic Friedmann equation ($H = 1/l_{IR}$, as mentioned above) for a suitable choice of $L_{IR}$. 
Therefore the questions that naturally arise are:
\begin{itemize}
 \item Is the holographic correspondence just confined to vacuum $R^2$ gravity or can be extended to other forms 
 of ``vacuum $F(R)$ gravity models'' ? Moreover what about the holographic correspondence for 
 non-vacuum F(R) models i.e the F(R) models along with matter fields ? 
\end{itemize}

We try to address these questions in the present paper.

\section{Holographic correspondence of F(R) gravity without matter fields}
First, we show that the holographic inflation with generalized infrared cut-off can reproduce the cosmological equations 
for higher curvature like $F(R)$ gravity (see \cite{nojiri} for review of F(R) gravity) in absence of matter fields. 
As for example, we determine the corresponding cut-off for three popular forms of F(R) gravity, in particular 
$F(R)=R + \alpha R^2$, $F(R)=R + \alpha R^3$ and $F(R)=e^{\beta R}$ respectively.

\begin{itemize}
 \item For quadratic curvature gravity, let us consider the generalized infrared cut-off ($L_{IR}$) as follows \cite{odintsov},

\begin{eqnarray}
 L^{(q)}_{IR} = -\frac{1}{6\alpha\dot{H}^2a^6} \int dt a^6\dot{H}
 \label{quadratic_cut-off}
\end{eqnarray}

with $\alpha$ be the model parameter and $q$ stands for quadratic curvature. 
Plugging back this expression into the holographic Friedmann equation $H = \frac{1}{L_{IR}}$ and 
simplifying a little bit yield the following differential equation of the Hubble parameter as, 

\begin{eqnarray}
 H^2 = 6\alpha \bigg[\dot{H}^2 - 2H\ddot{H} - 6\dot{H}H^2\bigg]
 \label{quadratic_equation}
\end{eqnarray}
With the information that the Ricci scalar in FRW geometry can be written as $R = 12H^2 + 6\dot{H}$, it can be easily shown that 
eqn.(\ref{quadratic_equation}) resembles with the cosmological equation for $F(R) = R + \alpha R^2$ gravity. Thus the holographic inflation with the 
cut-off $L^{(q)}_{IR}$ (see eqn.(\ref{quadratic_cut-off})) can reproduce the Starobinsky $R^2$ inflation \cite{43,44,47} which is known to lead to 
inflationary observables in a very good agreement with observations.\\

\item For cubic curvature gravity, starting with the following generalized infrared cut-off

\begin{eqnarray}
 &L^{(c)}_{IR}& = \nonumber\\
 &-&\frac{\int dt a^{15/2}\dot{H}}
 {36\bigg[2\alpha\dot{H}^3a^{15/2} - 4\alpha H\int dt a^{15/2}H\dot{H}\bigg(H^3 - 9H\dot{H} - 3\ddot{H}\bigg)\bigg]}\nonumber\\
 \label{cubic_cut-off}
\end{eqnarray}

and inserting this expression into $H = 1/L_{IR}$, one lands up with the following equation:
\begin{eqnarray}
 H^2&=&36\alpha \bigg[2\dot{H}^3 - 6H\dot{H}\ddot{H} - 15H^2\dot{H}^2 + 4H^6\nonumber\\
 &-&36H^4\dot{H} - 12H^3\ddot{H}\bigg]
 \label{cubic_equation}
\end{eqnarray}
with $\alpha$ be the model parameter. Eqn.(\ref{cubic_equation}) represents the Friedmann equation for $F(R) = R + \alpha R^3$ gravity. Thereby the 
cosmology of $R^3$ gravity can be realized from holographic origin with a specified infrared cut-off determined in eqn.(\ref{cubic_cut-off}). 
It is well known that $F(R) = R + \alpha R^3$ does not give a good inflation, in particular the theoretical values of $n_s$ and $r$ do not support 
the observable constraints from Planck 2018. However in the present context, our main concern is not about presenting a 
viable inflationary model, rather to show how the holographic cut-off can mimic the cosmological equations for various F(R) gravity 
in absence of matter fields. The corresponding cut-off for higher curvature gravity $with$ matter fields will be presented in later section. 
However it was shown earlier \cite{tp1} that in presence of matter field, in particular the second rank antisymmetric Kalb-Ramond field, 
the cubic gravity model is consistent with the Planck 2018 constraints (i.e $n_s = 0.9650 \pm 0.0066$ and $r < 0.07$).

\item For exponential F(R) gravity, the corresponding infrared cut-off is given by,

\begin{eqnarray}
 L^{(e)}_{IR} = \frac{\int dt a^4\big(1 - 6\beta\dot{H}^2/H^2\big)}
 {6\beta a^4\dot{H} + H\int dt a^4\bigg(\frac{1}{6\beta H^2} - \frac{\dot{H}}{H^2}\bigg)}
 \label{exponential_cut-off}
\end{eqnarray}
where $\beta$ be the model parameter. This expression of $L^{(e)}_{IR}$ along with the holographic equation $H = 1/L_{IR}$ lead to the following 
differential equation for $H$ :
\begin{eqnarray}
 H^2 = 6\beta \bigg[H\ddot{H} + 4H^2\dot{H}\bigg] + \frac{1}{6\beta}\big[1 - 6\beta\dot{H}\big]
 \label{exponential_equation}
\end{eqnarray}
which can be re-written in the form 
\begin{eqnarray}
 \frac{F(R)}{2} = 3(H^2 + \dot{H}) F'(R) - 18(4H^2\dot{H} + H\ddot{H})F''(R)
 \nonumber
\end{eqnarray}
with $F(R) = e^{\beta R}$. Therefore the holographic equation can mimic the cosmological equations of exponential F(R) gravity, thanks to 
the holographic cut-off in eqn.(\ref{exponential_cut-off}). 
Moreover unlike to cubic curvature gravity, the exponential F(R) model is known to be in good agreement with 
Planck constraints \cite{odintsov2}.\\
\end{itemize}
At this stage it deserves mentioning that in $F(R)$ gravity model $S = \frac{1}{2\kappa^2}\int d^4\sqrt{-g}F(R)$, 
the higher curvature term(s) act as $geometric$ matter field with stress tensor given by
\begin{eqnarray}
 T_{\mu\nu}^{(f)} = \frac{1}{\kappa^2}\bigg[\frac{1}{2}g_{\mu\nu}f(R) - R_{\mu\nu}f'(R) + \bigg(\nabla_{\mu}\nabla_{\nu}-g_{\mu\nu}\Box\bigg)f'(R)\bigg]
 \nonumber\\
 \label{new1}
\end{eqnarray}
where $f(R) = F(R) - R$. The authors of \cite{cappo2} showed that $T_{\mu\nu}^{(f)}$ describes a $perfect~fluid$ provided the following conditions-
\begin{eqnarray}
 R_{\mu\nu}&=&\frac{R-4\xi}{3}u_{\mu}u_{\nu} + \frac{R-\xi}{3}g_{\mu\nu}\label{con1}\\
 0&=&\nabla_{\mu}R + u_{\mu}u^{\nu}\nabla_{\nu}R\label{con2}\\
 \nabla_{\mu}u_{\nu}&=&\varphi\bigg(g_{\mu\nu} + u_{\mu}u_{\nu}\bigg)
 \label{con3}
\end{eqnarray}

are satisfied, where $\varphi$ is a scalar field, $u_{\mu}$ is a timelike vector field having $u_{\mu}u^{\mu}=-1$ and $\xi$ obeys the equation 
$R_{\mu\nu}u^{\nu} = \xi u_{\mu}$. 
With these three conditions, $T_{\mu\nu}^{(f)}$ acts as $T_{\mu\nu}^{(f)} = \big(\mu + p\big)u_{\mu}u_{\nu} + pg_{\mu\nu}$ (i.e as 
perfect fluid) with energy density and pressure given by
\begin{eqnarray}
\mu&=&\frac{1}{\kappa^2}\bigg[-\frac{1}{2}f(R) + \xi f'(R) - 3\varphi \dot{R}f''(R)\bigg]\nonumber\\
p&=&\frac{1}{\kappa^2}\bigg[\frac{1}{2}f(R) - \frac{R-\xi}{3}f'(R) + \big[3\varphi\dot{R} + \ddot{R}\big]f''(R)\nonumber\\
&+&\big(\dot{R}\big)^2f'''(R)
\label{new3}
\end{eqnarray}
respectively. However in a FRW spacetime, the conditions from eqn.(\ref{con1}) to (\ref{con3}) are assured for 
the choices : $\xi = -R_{00} = 3\big(\dot{H} + H^2\big)$, 
$\varphi = H$ and $u^{\mu} = (1,0,0,0)$. These choices of $\xi$ and $\varphi$ lead to the energy density ($\mu$) 
for quadratic, cubic and exponential F(R) models as
\begin{eqnarray}
 \frac{\kappa^2}{3}\mu^{(q)}&=&6\alpha \bigg[\dot{H}^2 - 2H\ddot{H} - 6\dot{H}H^2\bigg]\label{sim1}\\
 \frac{\kappa^2}{3}\mu^{(c)}&=&36\alpha \bigg[2\dot{H}^3 - 6H\dot{H}\ddot{H} - 15H^2\dot{H}^2 + 4H^6\nonumber\\
 &-&36H^4\dot{H} - 12H^3\ddot{H}\bigg]\label{sim2}\\
 \frac{\kappa^2}{3}\mu^{(e)}&=&6\beta \bigg[H\ddot{H} + 4H^2\dot{H}\bigg] + \frac{1}{6\beta}\big[1 - 6\beta\dot{H}\big]
 \label{sim3}
\end{eqnarray}

where the superscripts stand as : 'q' for quadratic, 'c' for cubic and 'e' for exponential F(R) model. 
It may be observed that right hand sides of eqns.(\ref{sim1}) to (\ref{sim3}) match with that of eqns.(\ref{quadratic_equation}), 
(\ref{cubic_equation}) and (\ref{exponential_equation}) respectively. Thus the F(R) models described in the earlier subsections can act 
as a perfect fluid, in particular a $geometric~perfect~fluid$ over the four dimensional FRW spacetime.\\
Till now we describe the generalized infrared cut-offs for various F(R) gravity models 
in $absence$ of matter fields (apart from geometric matter fields). 
However it is also important to investigate whether such holographic correspondence for higher curvature 
cosmological models are just confined to vacuum case or can be extended to the F(R) models along $with~matter~fields$. 
This is discussed in the following section.\\

\section{Holographic correspondence of F(R) gravity with matter fields}
\begin{itemize}
 \item Without loss of generality, we start by considering an Infrared cut-off of the form

\begin{eqnarray}
 L_{IR} = -\frac{1}{6\alpha\dot{H}^2a^6} \int dt a^6\dot{H} \bigg(1 - \frac{\kappa^2\rho_0}{a^{3(1+w)}H^2}\bigg)
 \label{quadratic_field_cut-off}
\end{eqnarray}

with $\rho_0$, $\alpha$ and $w$ be the model parameters. Inserting it into $H = 1/L_{IR}$ along with a little bit simplification, 
we land up with the following differential equation, 

\begin{eqnarray}
 H^2 = 6\alpha \bigg[\dot{H}^2 - 2H\ddot{H} - 6\dot{H}H^2\bigg] + \frac{\kappa^2\rho_0}{a^{3(1+w)}}
 \label{quadratic_field_equation}
\end{eqnarray}

Eqn.(\ref{quadratic_field_equation}) is actually a combination of two separate equations as,
\begin{eqnarray}
 \frac{F(R)}{2}&=&3(H^2 + \dot{H}) F'(R)\nonumber\\
 &-&18(4H^2\dot{H} + H\ddot{H})F''(R) + \rho_{mat}
 \label{1}
\end{eqnarray}
and
\begin{eqnarray}
 \dot{\rho} + 3H\rho (1+w) = 0
 \label{2}
\end{eqnarray}
respectively with $F(R) = R + \alpha R^2$. Therefore with the infrared cut-off determined in eqn.(\ref{quadratic_field_cut-off}), the holographic 
equation can reproduce the cosmological scenario of $R^2$ gravity along with matter field having equation of 
state parameter denoted by $w$ and $\rho_0$ is identified with the integration constant in solving of eqn.(\ref{2}). Moreover 
for $\rho_0 = 0$, the cut-off in eqn.(\ref{quadratic_field_cut-off}) matches with the expression in eqn.(\ref{quadratic_cut-off}), as expected.\\
Whether the fluid is scalar or vector or higher rank tensor field depends on the state parameter $w$ which in turn fixes 
the form of the cut-off according to eqn.(\ref{quadratic_field_cut-off}): as for example - \\
(1) a massless scalar field has $w=1$, since the energy density($\rho$) 
and pressure ($p$) of a massless scalar field is given by $\rho = \frac{1}{2}\dot{\Phi}^2$ and $p = \frac{1}{2}\dot{\Phi}^2$ respectively where 
$\Phi$ is the scalar field,\\ 
(2) the electromagnetic vector field has $w=1/3$ which comes from the fact that the trace of energy momentum tensor of 
electromagnetic field vanishes,\\
(3) the second rank antisymmetric Kalb-Ramond (KR) field has $w=1$ which gives its evolution as 
$\rho_{KR} = \rho_0/a^6$ i.e the energy density of KR field decreases with a faster rate in comparison to pressureless matter and radiation. This explains 
why the present universe is practically free from any signatures of KR field. However the KR field has considerable effects in early universe, 
in particular the presence of KR field enhances the value of tensor to scalar ratio with respect to F(R) gravity alone \cite{tp1,tp2}. 
Earlier it was shown \cite{tp1} that $R^2$ gravity with KR field gives a $good~inflation$ i.e supports an exit of inflation as well 
as the theoretical values of $n_s$ and $r$ 
are consistent with Planck 2018 constraints. For completeness, here we give the the simultaneous plots of $n_s$, $r$ for the specified range 
of the model parameters, see Figure[\ref{plot1}].\\

\begin{figure}[!h]
\begin{center}
 \centering
 \includegraphics[width=3.0in,height=2.5in]{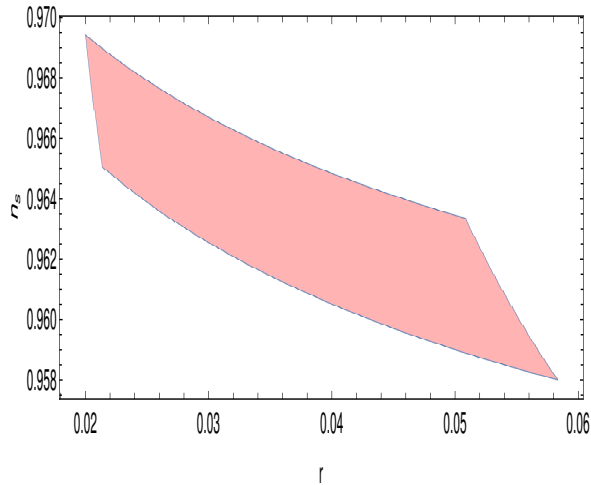}
 \caption{$n_s$ (along y axis) vs $r$ (along x axis) for $F(R)=R+\alpha R^2$ with Kalb-Ramond field within the parametric range 
 $0.003 \lesssim \kappa^2\rho_0{\alpha} \lesssim 0.004$}
 \label{plot1}
\end{center}
\end{figure}

However, observations based on Planck $2018$ put a constraint on $n_s$ and $r$ as 
 $n_s = 0.9650 \pm 0.0066$ and $r < 0.07$ (combining with BICEP2/Keck - Array) respectively. 
 Therefore, figure \ref{plot1} clearly shows that for $0.003 < \kappa^2\rho_0\alpha < 0.004$, 
 the theoretical values of $n_s$, $r$ (in the present context) match with the observational constraints. 
In addition, by the estimated values of $\kappa^2\rho_0\alpha$, the duration of inflation becomes 
 $10^{-12}$(Gev)$^{-1}$ if $\frac{1}{\sqrt{\alpha}}$ is separately taken as $10^{-5}$ (in Planckian unit) \cite{tp1}. Within this parametric regime, the 
 number of e-foldings (N) comes as $N = 56$ \cite{tp1}. These results are summarized in Table \ref{Table-1}.

\begin{table}[!h]
 \centering
  \begin{tabular}{|c| c|}
   \hline \hline
   Parameters & Estimated values\\
   \hline
   $n_s$ & $\simeq 0.9625$\\ 
   $r$ & $\simeq 0.03$\\
   Duration & $10^{-12}$(GeV)$^{-1}$\\
   $N$ & 56\\
   \hline
  \end{tabular}%
  \caption{Estimated values of various quantities for $\kappa^2\rho_0\alpha = 0.0035$ and $1/\sqrt{\alpha} = 10^{-5}$ (in Planckian unit) in 
  $R+\alpha R^2+$KR field model.}
  \label{Table-1}
 \end{table}

Thus the holographic inflation with the cut-off determined in eqn.(\ref{quadratic_field_cut-off}) (with $w=1$) reproduces $R^2$ gravity with 
second rank antisymmetric KR field, which is known to lead a good inflationary dynamics and also the observable parameters lie within the Planck 2018 
constraints.\\

\item In order to generate the cubic curvature gravity with matter fields from holographic principle, we consider the following Infrared cut-off,

\begin{eqnarray}
 &L_{IR}& =\nonumber\\
 &-&\frac{\int dt a^{15/2}\dot{H} \bigg(1 - \frac{\kappa^2h_0}{a^{3(1+w)}H^2}\bigg)}
 {36\bigg[2\alpha\dot{H}^3a^{15/2} - 4\alpha H\int dt a^{15/2}H\dot{H}\bigg(H^3-9H\dot{H}-3\ddot{H}\bigg)\bigg]}\nonumber\\
 \label{cubic_field_cut-off}
\end{eqnarray}
Plugging back this expression into $H = 1/L_{IR}$, one lands up with the following cosmological equation of Hubble parameter:
\begin{eqnarray}
 H^2&=&36\alpha \bigg[2\dot{H}^3 - 6H\dot{H}\ddot{H} - 15H^2\dot{H}^2 + 4H^6\nonumber\\
 &-&36H^4\dot{H} - 12H^3\ddot{H}\bigg] + \frac{\kappa^2]\rho_0}{a^{3(1+w)}}
 \label{cubic_field_equation}
\end{eqnarray}
Eqn.(\ref{cubic_field_equation}) resembles with the cosmological scenario of $R^3$ gravity model along with matter fields having $w$ be 
the equation of state parameter. Therefore the cut-off determined in eqn.(\ref{cubic_field_cut-off}) along with the holographic equation 
$H = 1/L_{IR}$ encodes the cosmological informations of $R+\alpha R^3$ gravity in presence of matter fields.\\
As mentioned earlier, unlike to vacuum $R^3$ gravity, the $F(R)=R+\alpha R^3$ model along with Kalb-Ramond field (having $w=1$) leads to 
the inflationary parameters which are in good agreement with the latest Planck constraints, as shown in \cite{tp1}. However for completeness, here 
we present the simultaneous plots of $n_s$, $r$ in Figure[\ref{plot2}] which clearly demonstrates that the theoretical values of 
$n_s$ and $r$ fit well with the Planck constraints.\\

\begin{figure}[!h]
\begin{center}
 \centering
 \includegraphics[width=2.5in,height=2.5in]{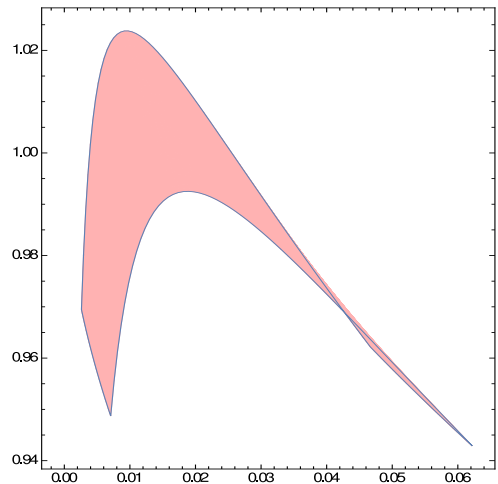}
 \caption{$n_s$ (along y axis) vs $r$ (along x axis) for $F(R)=R+\alpha R^3$ with Kalb-Ramond field within the parametric range 
 $0.03 \lesssim \kappa^2\rho_0\sqrt{\alpha} \lesssim 0.3$}
 \label{plot2}
\end{center}
\end{figure}

\item For exponential F(R) gravity $+$ matter fields, we take the infrared cut-off as follows:

\begin{eqnarray}
 L_{IR} = \frac{\int dt a^4\bigg(1 - 6\beta\dot{H}^2/H^2 - \frac{\kappa^2\rho_0}{a^{3(1+w)}H^2}\bigg)}
 {6\beta a^4\dot{H} + H\int dt a^4\bigg(\frac{1}{6\beta H^2} - \frac{\dot{H}}{H^2}\bigg)}
 \label{exponential_field_cut-off}
\end{eqnarray}
This expression along with the equation $H = 1/L_{IR}$ immediately lead to the following differential equation for the Hubble parameter, 
\begin{eqnarray}
 H^2 = 6\beta \bigg[H\ddot{H} + 4H^2\dot{H}\bigg] + \frac{1}{6\beta}\big[1 - 6\beta\dot{H}\big] + \frac{\kappa^2h_0}{a^{3(1+w)}}\nonumber\\
 \label{exponential_field_equation}
\end{eqnarray}

Eqn.(\ref{exponential_field_equation}) represents the cosmological scenario of $F(R) = e^{\beta R}$ along with matter fields if $w$ is 
identified with the equation of state parameter of the matter field. The inflationary phenomenology of vacuum exponential gravity is already presented 
in \cite{odintsov2}. However here we briefly discuss the phenomenology of the $F(R)=e^{\beta R}$ model in presence of the second 
rank antisymmetric Kalb-Ramond field, recall, KR field has $w=1$. 
The spectral index and tensor to scalar ratio in the present context are defined as follows \cite{tp1,new1,new2,tp2}:
 \begin{eqnarray}
  n_s = \big[1 - 4\epsilon_F - 2\epsilon_2 + 2\epsilon_3 - 2\epsilon_4\big]\bigg|_{\tau_0}\ ,
  \label{spectral index1}
 \end{eqnarray}
 and
 \begin{eqnarray}
  r = 8\kappa^2 \frac{\varTheta}{F'(R)}\bigg|_{\tau_0}\ .
  \label{ratio1}
 \end{eqnarray}
Here the slow roll parameters ($\epsilon_F$, $\epsilon_2$, $\epsilon_3$, $\epsilon_4$) are defined by the following expressions,
\begin{eqnarray}
 \epsilon_F&=&-\frac{1}{H^2} \frac{dH}{d\tau}\ ,~~~~~~~~~~\epsilon_2 = \frac{1}{2\rho_{KR}H} \frac{d\rho_{KR}}{d\tau}\ ,\nonumber\\
 \epsilon_3&=&\frac{1}{2F'(R)H} \frac{dF'(R)}{d\tau}\ ,~~~~~~~~~~\epsilon_4 = \frac{1}{2EH} \frac{dE}{d\tau}\ ,
 \label{various slow roll parameters}
\end{eqnarray}
where $\varTheta$ and $E$ are given by,
\begin{eqnarray}
 \varTheta = \frac{\rho_{KR}}{F'(R)H^2}\bigg[F'(R) + \frac{3}{2\kappa^2\rho_{KR}}\bigg(\frac{d}{d\tau}F'(R)\bigg)^2\bigg]\ ,
 \label{vartheta}
\end{eqnarray}
and
\begin{eqnarray}
 E = \frac{\varTheta F'(R)H^2}{\rho_{KR}}\ ,
 \label{E}
\end{eqnarray}
with $\rho_{KR}$ ($= H_{123}H^{123}$) being the energy density of the KR field. By using the field equations, one can simplify 
the expressions of $n_s$ and $r$ to yield

\begin{eqnarray}
 n_s = 1 - 2\frac{\epsilon_F'}{\epsilon_F H} 
 + \frac{\kappa^2\rho_{KR}}{6F'(R)\epsilon_F^2 H^3} \bigg[-6H - \frac{H'}{H}\bigg]\ .
 \label{spectral index intermediate}
\end{eqnarray}
and
\begin{eqnarray}
 r = 8\kappa^2 \frac{\rho_{KR}}{H^2 F'(R)} + 48 \bigg(\frac{1}{2H F'(R)}\frac{dF'(R)}{d\tau}\bigg)^2\bigg|_{\tau_0}\ .
 \label{fifth 1}
\end{eqnarray}
respectively. Having these expressions in hand along with $F(R)=e^{\beta R}$, we give the simultaneous plot of $n_s$, $r$ in Figure[\ref{plot3}]  
which reveals that the theoretical values of $n_s$ and $r$ in the present context lie within the Planck 2018 constraints.

\begin{figure}[!h]
\begin{center}
 \centering
 \includegraphics[width=2.5in,height=2.7in]{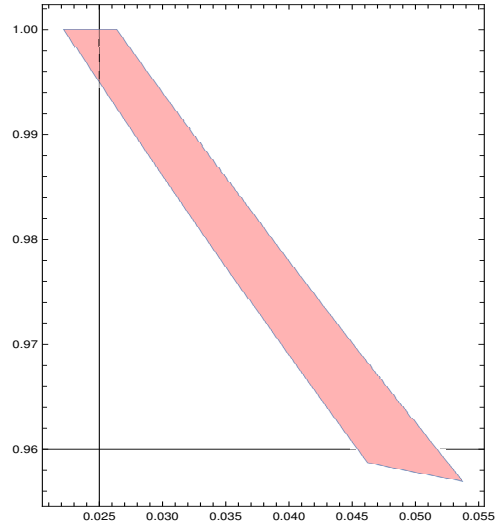}
 \caption{$n_s$ (along y axis) vs $r$ (along x axis) for $F(R)=e^{\beta R}$ with Kalb-Ramond field.}
 \label{plot3}
\end{center}
\end{figure}
\end{itemize}

Hence we can argue that the generalized infrared cut-off obtained in eqn.(\ref{exponential_field_cut-off}) can reconstruct the 
cosmological field equations for exponential F(R) gravity in presence of Kalb-Ramond field, which gives the inflationary parameters 
to be in good agreement with Planck constraints, as confirmed by Fig.[\ref{plot3}].\\
The above subsections clearly demonstrate that the holographic correspondence of higher curvature cosmological models 
are not only confined in vacuum but also can be extended to the F(R) models along with the matter fields.

\section{Holographic correspondence of F(T) cosmology}
We extend our discussion of holographic correspondence to the generalized teleparallel cosmology i.e F(T) cosmology \cite{cappo,saridakis,bamba}. 
The teleparallel gravity is described by the Weitzenbock connection which is curvature-free, unlike to Einstein's General Relativity. 
This new connection is determined by two dynamical variables, namely the tetrads and the spin connection. In the current work, we consider the 
flat FRW metric i.e
\begin{eqnarray}
 ds^2 = -dt^2 + a^2(t)\big[dx^2 + dy^2 + dz^2\big]
 \label{t1}
\end{eqnarray}
where $a(t)$ is the scale factor of the Universe. The natural choice of tetrad for this metric is 
\begin{eqnarray}
e^a_{\mu} = diag\big(1,a(t),a(t),a(t)\big)
\label{t2}
\end{eqnarray}
This form of the FRW metric is very advantageous because its spin connection vanishes \cite{saridakis}, and so 
no extra contribution is needed in the F(T) field equations. Analogous to Ricci scalar in GR, one can define torsion scalar in teleparallel 
gravity, which plays the role of Lagrangian for the teleparallel equivalent of general relativity (TEGR) case. For the aforementioned tetrad 
in eqn.(\ref{t2}), the torsion scalar turns out to be $T = -6H^2$, with $H$ being the Hubble parameter.\\ 
Following the same reasoning as F(R) gravity, the action of TEGR can be generalized to  
\begin{eqnarray}
 S = \frac{1}{2\kappa^2} \int d^4x |e|~F(T)
 \label{t3}
\end{eqnarray}
where $e = det~e^a_{\mu}$ and $F(T)$ is an analytic function of $T$. Variation of the action with respect to the tetrad along with the consideration 
of flat FRW spacetime lead to the following equation:
\begin{eqnarray}
 H^2 = -\frac{\big(F(T) - T\big)}{6} - 2H^2~\frac{dF}{dT}
 \label{t4}
\end{eqnarray}
For cosmological interest, we consider a power law model of the following form \cite{cappo,190}
\begin{eqnarray}
 F(T) = T - \alpha\big(-T\big)^p
 \label{t5}
\end{eqnarray}
Earlier it was shown that the F(T) model in eqn.(\ref{t5}) (along with suitable initial conditions) allow the universe to evolve from an 
initial phase of radiation domination to a cosmic acceleration at late times for $p \neq 1$ \cite{cappo}. By substituting the above form 
of F(T) into the Friedmann equation (\ref{t4}) and after some simplification, we get a non-zero constant expression of the Hubble parameter as follows,
\begin{eqnarray}
 H^2 = \frac{1}{6}\bigg[\frac{3}{\alpha(1-2p)}\bigg]^{1/(p-1)}
 \label{t6}
\end{eqnarray}
However eqn.(\ref{t6}) can be reconstructed from the holographic Friedmann equation (i.e $H = 1/L_{IR}$) with the infrared cut-off given by,

\begin{eqnarray}
 L_{IR} = \sqrt{6}~\bigg[\frac{\alpha(1-2p)}{3}\bigg]^{\frac{1}{2(p-1)}}
 \label{t7}
\end{eqnarray}

Therefore the power law F(T) model shown in eqn.(\ref{t5}) has a holographic correspondence with the infrared cut-off determined in eqn.(\ref{t7}). 
Similarly, the cosmological field equations for other F(T) models can also be reproduced from $H = 1/L_{IR}$ with suitable infrared cut-off, for example :
\begin{itemize}
 \item $L_{IR} = \sqrt{6}~\bigg[\frac{\beta(1-2q)}{2}\bigg]^{\frac{1}{2(q-1)}}$ for $F(T) = - \beta\big(-T\big)^q$
\end{itemize}
and
\begin{itemize}
 \item $L_{IR} = \frac{3}{\sqrt{\Lambda}}$ for $F(T) = T - 2\Lambda$
\end{itemize}
respectively. The above two forms of F(T) gravity help to describe the early universe inflation as demonstrated in \cite{myrzakulov_new}.\\
Thereby we can argue that similar to F(R) gravity, the F(T) model(s) may also be realized from holographic origin.

\section{Conclusion}
In this paper, we apply the holographic principle at the early universe, obtaining an inflation realization of holographic origin. 
We construct generalizations of holographic inflation which are based on extended Infrared cut-offs. Under these considerations, it was 
shown earlier \cite{odintsov} that the scenario of holographic inflation, under the generalized Infrared cutoff can reproduce 
Starobinsky $R^2$ inflation. However in the present context, we show that with the help of generalized holographic cut-off, one can go 
beyond the Starobinsky inflation and can reconstruct the cosmological field equations of other F(R) models also, in particular the cubic 
and exponential F(R) gravity in absence of matter fields. Moreover we show that such holographic correspondence is not only confined 
to the vacuum F(R) gravity, but also can be extended to the cosmological F(R) models along with matter fields. We explicitly determine 
the infrared cut-offs for various F(R) models (such as quadratic, cubic and exponential F(R) gravity) along with matter fields having a constant 
equation of state (e.o.s) parameter : massless scalar field, electromagnetic vector field, second rank antisymmetric Kalb-Ramond field etc. belong from 
this class (i.e constant e.o.s parameter) of matter field. In presence of matter field, in particular the Kalb-Ramond field, 
all the considered F(R) models are known 
to lead the inflationary parameters which fit well with the Planck 2018 constraints. We further extend our discussion to F(T) cosmology 
and show that similar to F(R) gravity, the F(T) model(s) may also be realized from holographic origin. Thus the holographic inflation proves 
to have a very interesting phenomenology, which makes it a good candidate for the description of the early universe.\\
However it is important to consider other generalized Infrared cutoffs in order to obtain a correspondence with 
other geometrical inflationary models, such as Gauss-Bonnet,  $F(G)$ inflation \cite{47,nojiri} 
and moreover the models containing the matter fields 
with variable equation of state parameter. These are expected to investigate in a future work.

\end{document}